\documentclass[twocolumn,showpacs,preprintnumbers,amsmath,amssymb]{revtex4}

\usepackage{graphicx}
\usepackage{dcolumn}
\usepackage{bm}

\begin{document}

\preprint{APS/123-QED}


\title{Radiation induced oscillations of the Hall resistivity 
in two-dimensional electron systems} 
\author{ 
V.~Ryzhii}
\affiliation{Computer Solid State Physics Laboratory, University of Aizu,
Aizu-Wakamatsu 965-8580, Japan}

\date{\today}

\begin{abstract}
We consider the effect of microwave radiation 
on  the Hall 
resistivity in two-dimension electron systems.
It is shown that    photon-assisted
impurity scattering of electrons can result in oscillatory
dependences 
of both dissipative and Hall components of the conductivity
and resistivity
tensors on the ratio of radiation frequency to cyclotron frequency. 
The Hall resistivity can include a  component induced by 
microwave radiation
which is an even function of the magnetic field.
The phase of the dissipative resistivity oscillations
and  the polarization dependence of their amplitude are compared
with those of the Hall
resistivity oscillations.
The developed model can clarify the results of
recent experimental observations of the radiation induced Hall effect.
\end{abstract}

\pacs{73.40.-c, 78.67.-n, 73.43.-f}


\maketitle

The oscillatory dependence of the conductivity
of  two-dimensional electron systems (2DES's) under microwave radiation
in a magnetic field on the ratio of radiation frequency $\Omega$
to cyclotron frequency $\Omega_c$
 has been predicted
more than three decades ago by the author~\cite{1} 
(see also Ref.~\cite{2}). As shown~\cite{1,2}, the variation
of the dissipative conductivity caused by irradiation (microwave 
photoconductivity) is negative 
if $\Omega$ somewhat exceeds $\Lambda\Omega_c$,
where $\Lambda = 1, 2, 3,..$. In this situation, the net
dissipative conductivity can become negative if the incident
microwave power is sufficiently high 
(effect of absolute negative conductivity or ANC). 
The oscillations of
the dissipative components of the conductivity tensor
$\sigma_{xx}$ and $\sigma_{yy}$
considered in Refs.~\cite{1,2}  are associated with  photon-assisted
impurity scattering of electrons accompanied by their transitions
between the Landau levels (LL's). 
As shown in Refs.~\cite{3,4,5,6},  photon-assisted
impurity scattering can result in variations of the Hall
conductivity components $\sigma_{xy}$ and $\sigma_{yx}$
(with $\sigma_{xy} \neq - \sigma_{yx}$ due to an even dependence
of such variations on the magnetic field). 
The oscillatory dependences of the dissipative
conductivity in a 2DES as a function of $\Omega/\Omega_c$
were observed experimentally by Zudov {\it et al.}~\cite{7}
and Ye {\it et al.}~\cite{8}.
Shortly thereafter Mani {\it et al.}~\cite{9} and
 Zudov {\it et al.}~\cite{10} observed vanishing electrical
resistance $R_{xx}$ ($R_{xx} \propto  \rho_{xx} =
\sigma_{xx}/(\sigma_{xx}^2 + \sigma_{xy}^2)$
with increasing microwave power and formation
of the so-called zero-resistance states. This effect
is  attributed to reaching of the  threshold of
ANC~\cite{5,6,11,12,13} associated with the mechanism
considered in Refs.~\cite{1,2}.
The feasibility of the latter is 
owing to high electron mobility
in the 2DES's studied, so that  
the cyclotron resonance and its harmonics can 
be observed at rather small magnetic fields and,
hence, at microwave frequencies about 100~GHz and less.
Since the microwave power  required for ANC is approximately
proportional to $\Omega^3$,
 microwave sources with moderate powers can be used 
at moderate magnetic fields.
Further experimental results related to
the oscillations of microwave photoconductivity and
their consequences below and above
the ANC threshold
are presented in Refs.~\cite{14,15,16,17,18,19}.
In particular, Studenikin, {\it et al.}~\cite{18}
and Mani, {\it et al.}~\cite{19} demonstrated
the oscillations of the Hall resistance $R_{xy}$
which is related to the conductivity components as
$R_{xy} \propto  \rho_{xy} = \sigma_{xy}/(\sigma_{xx}^2 + \sigma_{xy}^2)$.
One can assume that the origin of these oscillations 
is analogous to that discussed previously~\cite{3,4,5,6}.

In this paper, we calculate the dependences of $\rho_{xx}$
and 
$\rho_{xy}$ vs
$\Omega_c/\Omega$ using a model invoking 
 photon-assisted impurity scattering of electrons as
the main mechanism determining  the microwave photoconductivity. 
The model under consideration
 generalizes that of
Refs.~\cite{3,4} in line with our recent papers~\cite{20,21}.
In principle, the concept based on photon-assisted impurity
scattering as the main mechanism responsible for 
the observed microwave photoconductivity explains virtually
all features of the effect, although it is prematurely
to draw the final conclusions yet. Namely, the positions of
the photoconductivity zeros, maxima, and minima~\cite{1,2,5,6},
saturation of the oscillation magnitude with increasing
microwave power and, hence, relatively large magnitude
of maxima and minima corresponding to 
the cyclotron resonance harmonics~\cite{20,22}. A significant
sensitivity of the amplitude
of the microwave photoconductivity oscillations to the temperature
can be attributed to the contribution of  photon-assisted acoustic
phonon scattering~\cite{21,23} and an increase of the LL broadening 
due to increasing dependence  of the electron-electron 
and acoustic scattering
on the temperature (see, for example, Refs.~\cite{24} and ~\cite{25},
respectively). The experimental observation of ANC~\cite{26} 
associated with photon-assisted electron transitions between
spatially separated states in the neighboring quantum wells
(microwave hopping photoconductivity associated with 
similar transitions between impurity states
was analyzed in Ref.~\cite{27} )
is a strong argument in favor of the dynamical mechanism
of ANC in 2DES's.
 Generally, the effect of ANC in a 2DES~\cite{28,29} and 
in a 3DES~\cite{30,31,32}
subjected to  a magnetic field
can also be associated with a deviation of the electron
distribution function from equilibrium one. 
Dorozhkin~\cite{17} proposed
a model based on the assumption that the electron distribution
function is nonmonotonic with inversion population of the
states near the center of LL's caused by the absorption
of microwave radiation. In this model, the electron Larmor orbit centers 
hop along the dc electric field 
(contributing to the dissipative conductivity) both immediately
absorbing microwave photons (absorption due to the photon-assisted
impurity scattering) and owing to nonradiative impurity scattering
upon excitation. If the maxima of the electron density of states
are shifted with respect to the maxima of the distribution function,
impurity scattering can, as shown by Elesin~\cite{30},
 lead to ANC. 
The main problem, however, is to realize such nontrivial
electron distributions.
Recently, Dmitriev, {\it et al.}~\cite{33}
developed a similar model applied to 2DES's irradiated
with microwaves in which, as assumed,
the electron distribution function can be oscillatory. 
In this regard, the consideration of the effect of
microwave radiation on the Hall conductivity is crucial to clarify 
genuine origin of the experimental results~\cite{7,8,9,10,14,15,16,17,18,19}.

The  probability of the electron transition between
the $(N, k_x, k_y)$ and $(N^{\prime}, k_x + q_x, k_y + q_y)$ 
states
in the presence of the net dc electric field ${\bf E} = (E,0,0)$
(including both the applied and Hall components)
perpendicular to the magnetic field ${\bf H} = (0,0,H)$
and the ac microwave field 
${\bf E}_{\Omega} = ({\cal E}e_x, {\cal E}e_y, 0)$
polarized in the 2DES plane ($e_x$ and $e_y$ are the components
of the microwave field complex polarization vector),
is given by  the following formula
obtained on the base of the interaction representation
of the operator of current via solutions of
the classical equations of electron motion~\cite{3,4,20}:

$$
W_{N, k_x, k_y; N^{\prime}, k_x + q_x, k_y + q_y}
= \frac{2\pi{\cal N}_i}{\hbar}\sum_{M}
|V_q|^2|Q_{N,N^{\prime}}(L^2q^2/2)|^2
$$ 
\begin{equation}\label{eq1}
\times J_M^2(\xi_{\Omega}(q_x, q_y ))\,
\delta[M\hbar\Omega + (N - N^{\prime})\hbar\Omega_c + eEL^2q_y].
\end{equation}
Here $N$ is the LL index, $k_x$ and $k_y$ are 
the electron quantum numbers,
$q_x$ and  $q_y$ are their variations due to
photon-assisted impurity scattering, $q = \sqrt{q_x^2 + q_y^2}$, 
$e = |e|$ is the electron charge, 
$\hbar$ is the Planck constant,
$M$ is the number
of absorbed or emitted  real microwave photons,  
${\cal N}_i$ is the impurity concentration,  $V_q$ is the matrix
element of the electron-impurity interaction, and
$Q_{N,N^{\prime}}(\eta) \propto  
L_N^{(N^{\prime} - N)}(\eta)\exp(-\eta/2)$,
where $L_N^{{\Lambda}}(\eta)$ is the Laguerre polynomial.
The LL form-factor is determined by the function
$\delta$ characterized by the LL broadening $\Gamma$. 
The effect of microwave radiation is reduced to the inclusion
of the energy $M\hbar\Omega$ of really absorbed 
or emitted $M$ photons 
in the argument 
of the function $\delta$ in Eq.~(1) and  the appearance of the Bessel
functions $J_M(\xi_{\Omega}(q_x, q_y ))$ ~\cite{3,4}.
Here  
$\xi_{\Omega}(q_x, q_y) = \displaystyle\frac{e{\cal E}}{m}
\frac{|q_xe_x + q_ye_y - i(\Omega_c/\Omega)(q_xe_y - q_ye_x)|}
{|\Omega_c^2 - \Omega(\Omega + i\Gamma )|}$,
is proportional to
 the amplitude of the electron orbit center oscillation
in the ac microwave field 
(limited by the dephasing of the electron motion),
where   $m$ is the electron effective mass. The quantity 
$\xi_{\Omega}(q_x, q_y)$ depends on
the polarization properties of the incident radiation.
At sufficiently high microwave power,
the amplitude of the electron orbit center oscillation
in the microwave field can become about the quantum Larmor radius $L$.

In the absence of microwave irradiation
at moderate dc electric fields $E \ll E_c =  \hbar\Omega_c/eL$,
the probability of inter-LL transitions is very small.
In this case, such transitions
 provide an exponentially small contribution
to the electron dissipative
conductivity~\cite{34}. 
At $E \lesssim E_b = \hbar\Gamma/eL$, 
the electron dissipative
conductivity  is due to electron scattering processes
associated with the intra-LL transitions~\cite{35}.
When a 2DES is irradiated with microwaves,
the variation of the dissipative conductivity
(microwave photoconductivity) can be associated with both
alteration of the intra-LL scattering (that can, in particular, 
result in the effect
of dynamic localization~\cite{26}) and appearance of
the photon-assisted transitions between LL's.
Considering Eq.~(1), the contribution
to the dissipative current $\Delta j_D$  (along
the axis $x$) associated with 
the photon-induced inter-LL  transitions
and  the variation
of 
the Hall current due to the  scattering processes in question
$\Delta j_H$ (along
the axis $y$) can be presented as
\begin{widetext}
\begin{equation}\label{eq2}
\Delta j_D = \frac{e{\cal N}_i}{\hbar} \sum_{M,N,N^{\prime} }
f_N(1 - f_{N^{\prime}})
\int d^2{\bf q}\,
q_y|V_q|^2|Q_{N,N^{\prime}}(L^2q^2/2)|^2
J_M^2(\xi_{\Omega}(q_x, q_y ))\,
\delta[M\hbar\Omega + (N - N^{\prime})\hbar\Omega_c + eEL^2q_y]\},
\end{equation}
\begin{equation}\label{eq3}
\Delta j_H = \frac{e{\cal N}_i}{\hbar} \sum_{M,N,N^{\prime} }
f_N(1 - f_{N^{\prime}})
\int d^2{\bf q}\,
q_x|V_q|^2|Q_{N,N^{\prime}}(L^2q^2/2)|^2
J_M^2(\xi_{\Omega}(q_x, q_y ))\,
\delta[M\hbar\Omega + (N - N^{\prime})\hbar\Omega_c + eEL^2q_y]\}.
\end{equation}
\end{widetext}
Here,
$f_N$ is the filling factor of the $N$th Landau level given
by the Fermi distribution function.
Assuming for definiteness that the LL density of states
is Lorentzian with the LL-broadening $\Gamma < \Omega, \Omega_c$,
at sufficiently small dc electric fields $E \ll E_b < E_c$,
and ac microwave fields $\xi_{\Omega}(q_x, q_y) \ll 1$,
for the quantities  $\Delta\sigma_D = j_D/E$ 
and $\Delta \sigma_H = \Delta j_H/E$ from Eqs.~(2) and (3), 
substituting the integration over $d^2{\bf q}$ for the integration
over $qdq\,d\theta$, where $\sin\theta = q_y/q$, one can 
obtain
\begin{equation}\label{eq4}
\Delta \sigma_D \propto (1 - e^{-\hbar\Omega/T})
\int_0^{2\pi}d\theta\sin^2\theta\, C(\theta,\Omega,\Omega_c),
\end{equation}
\begin{equation}\label{eq5}
\Delta\sigma_H \propto (1 - e^{-\hbar\Omega/T})
\int_0^{2\pi}d\theta\sin\theta\cos\theta\, C(\theta,\Omega,\Omega_c),
\end{equation}
where
\begin{equation}
C(P,\theta, \Omega, \Omega_c) 
= {\cal N}_iL \int dq q^2|V_q|^2
\exp(-l^2q^2/2)\nonumber
\end{equation}
\begin{equation}\label{eq6}
\times \sum_{M,\Lambda > 0}
\frac{J_M^2(Lq\sqrt{Pf}\Pi)
\Theta_{\Lambda}\Gamma(\Lambda\Omega_c - M\Omega )}
{[(\Lambda\Omega_c - M\Omega )^2 + \Gamma^2]^2}.
\end{equation}
Here $T$ is the temperature, $\Theta_{\Lambda} = \sum_Nf_N(1 - f_{N + \Lambda})/\sqrt{N}$, $P \propto c{\cal E}^2$ is the microwave power
normalized by  $p_{\Omega} = (m\Omega^3/2\pi\alpha)$ (where
$\alpha = e^2/\hbar c \simeq 1/137$ and $c$ is the speed of light),
$$
f(\Omega/\Omega_c) = \frac{(\Omega/\Omega_c)[1 + (\Omega/\Omega_c)^2]}
{[1 - (\Omega/\Omega_c)^2]^2 + (\Gamma/\Omega_c)^2(\Omega/\Omega_c)^2},
$$
and 
$\Pi(\theta) = \sqrt{2/[1 + (\Omega_c/\Omega)^2]}|e_x\cos\theta
+ e_y\sin\theta - 
i(\Omega_c/\Omega)^2(e_y\cos\theta - e_x\sin\theta)|$.
In deriving Eq.~(6), 
it is  assumed that many LL's are occupied,
so only the terms corresponding to the transitions
between upper LL's are important, and, therefore,
the  matrix elements in the  quasi-classical  limit can be used.   

Formulas~(4) - (6) describe 
the dissipative and Hall microwave photoconductivities as functions
of the normalized  power  of microwave radiation, 
its frequency and polarization, as well as the cyclotron frequency.
In particular, these formulas yield oscillatory dependences
of $\Delta \sigma_D$ and $\Delta \sigma_H$ on $\Omega/\Omega_c$
and a nonlinear  dependence of the amplitude  on $P$.
At $\Omega/\Omega_c \gtrsim \Lambda$, the microwave
dissipative photoconductivity $\Delta \sigma_D < 0$.
In this case, the net dissipative conductivity can be also negative
if the microwave power is 
sufficiently large. 
In the vicinity of the cyclotron resonance ($\Omega \sim \Omega_c$)
at moderate microwave powers $P \lesssim min\,f^{-1} \sim (\Gamma/\Omega_c)^2$,
one can take into account that $J_1^2(Lq\sqrt{Pf} 
\Pi) \sim q^2Pf\Pi^2$
and obtain
\begin{equation}\label{eq7}
C(P,\theta, \Omega, \Omega_c) 
\propto P\Pi^2(\theta)
\frac{\Gamma\, f(\Omega/\Omega_c)(\Omega_c - \Omega )}
{[(\Omega_c - \Omega )^2 + \Gamma^2]^2}S_i,
\end{equation}
where $S_i = {\cal N}_i\int_0^{\infty} dqq^4 |V_q|^2\exp(- L^2q^2/2)$.
Using Eqs.~(4), (5), and (7), we arrive at 
\begin{equation}\label{eq8}
\Delta \sigma_D \simeq P{\overline\Delta\sigma}\Phi_D, \qquad
\Delta \sigma_H \simeq P{\overline\Delta\sigma}\Phi_H, 
\end{equation}
with
\begin{equation}\label{eq9}
{\overline\Delta\sigma} \propto (1 - e^{-\hbar\Omega/T})
\frac{\Gamma\,f(\Omega/\Omega_c)(\Omega_c - \Omega )}
{[(\Omega_c - \Omega )^2 + \Gamma^2]^2}S_i,
\end{equation}

\begin{equation}\label{eq10}
\Phi_D = 
\frac{1}{2\pi}
\int_0^{2\pi}d\theta\sin^2\theta \Pi^2(\theta),
\end{equation}
\begin{equation}\label{eq11}
\Phi_H = \frac{1}{2\pi}\int_0^{2\pi}d\theta
\cos\theta\sin\theta \Pi^2(\theta).
\end{equation}
In the case of circular polarization, $\Pi(\varphi) = 1$ 
at any ratio $\Omega/\Omega_c$ and, hence,
$\Phi_D = 1/2 $ and  $\Phi_H~=~0$. 
At a linear polarization of the ac microwave field,
introducing the angle between the directions of
the dc and ac fields $\varphi$ so that $e_x = \cos\varphi$
and $e_y = \sin\varphi$, we obtain 
\begin{equation}\label{eq12}
\Phi_D = 
\displaystyle\frac{1}{4} + \frac{\Omega^2  + (\Omega_c^2 - \Omega^2)\sin^2\varphi}
{2(\Omega_c^2 + \Omega^2)},
\end{equation}
\begin{equation}\label{eq13}
\Phi_H~=~\displaystyle\frac{1}{4}\biggl(\frac{\Omega_c^2 - \Omega^2}
{\Omega_c^2 + \Omega^2}\biggr)\sin2\varphi.
\end{equation}
In the vicinity of the cyclotron resonance, Eqs.~(10) and (11)
yield
$\Phi_D \simeq 3/4$ and $\Phi_H
\simeq (\Omega_c^2 - \Omega^2)\sin2\varphi/8\Omega_c^2$.
The obtained polarization dependences are consistent with 
the general symmetry properties of the photoconductivity 
tensor~\cite{36} as well as with the previous calculations~\cite{3,4,6,37}.
In particular, as follows from Eqs.~(8) and (12), 
$\Delta \sigma_D$ exhibits larger
amplitude
of oscillations 
at $\varphi = \pi/2$ than at $\varphi = 0$ in agreement with
the results of Ref.~\cite{6}.
Equations~(8), (12), and (13) show that
the ratio of $\Delta \sigma_H$ and $\Delta \sigma_D$
at their maxima and minima corresponding to a cyclotron resonance harmonic
is larger that this ration at the maximum and minimum in the vicinity
of the cyclotron resonance,
where $\Phi_H \sim (\Gamma/\Omega_c)\sin2\varphi$. 

\begin{figure}[t]
\begin{center}
\includegraphics[width=6.5cm]{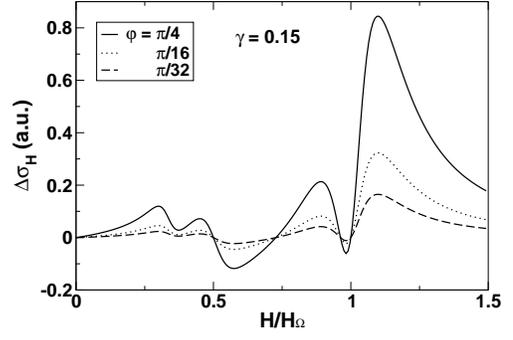}
\end{center}
\caption{Magnetic field dependencies of 
the Hall 
photoconductivity at different polarization angles.}
\end{figure}

\begin{figure}[t]
\begin{center}
\includegraphics[width=6.5cm]{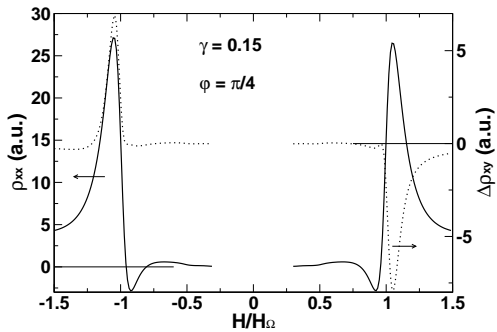}
\end{center}
\caption{Magnetic field dependencies of the dissipative (solid curves) 
and Hall (dotted curves)
resistivities at $\varphi = \pi/4$.}
\end{figure}

The conductivity tensor components are related to
$\Delta\sigma_D$ and $\Delta \sigma_H$
as $\sigma_{xx}
= \sigma_{yy}= \sigma_D + \Delta\sigma_D$, 
$\Delta \sigma_{xy} 
= \sigma_H + \Delta \sigma_H$,
and $\Delta\sigma_{yx} =  - \sigma_H + \Delta \sigma_H$,
where  $\sigma_D$ and $\sigma_H$ are the the dissipative
and Hall conductivities
without microwave radiation.
As follows from the above results, both $\Delta \sigma_D$
and $\Delta \sigma_H$ do not vary when the magnetic field
is reversed.
This implies
that in the situation under consideration, 
$\sigma_{xy}$
and $- \sigma_{yx}$ are not equal
to each other.
Expressing the components of the conductivity tensor
via $\sigma_D$, $\sigma_H$, $\Delta\sigma_D$, and $\Delta\sigma_H$,
taking into account that
$\sigma_H \gg \sigma_D, \Delta\sigma_D, \Delta\sigma_H$,
and using Eq.~(8),
the variation  of the Hall resistivity under the influence
of microwave radiation  can presented
in the following form:

\begin{equation}
\Delta\rho_{xy} \simeq -\frac{\Delta\sigma_H}{\sigma_H^2}
- \frac{(2\sigma_D  + \Delta\sigma_D)\Delta\sigma_D}{\sigma_H^3}\nonumber
\end{equation}
\begin{equation}\label{eq14}
\simeq - P\frac{{\overline\Delta\sigma}}{\sigma_H^2}
\biggl[\Phi_H
+ \frac{(2 \sigma_D + P{\overline\Delta\sigma}\Phi_D)\Phi_D}{\sigma_H}\biggr].
\end{equation}
Equation~(14) qualitatively describes the following
 effects:
(a) a shift of the $\rho_{xy}$ vs $H$ dependence (presence of a component
which is an odd function of the magnetic field)
stimulated by microwave radiation, (b)
antisymmetric oscillations of $\Delta\rho_{xy}$
with varying magnetic field,  and (c) markedly
different span of $\Delta\rho_{xy}$ at $H \gtrsim H_{\Omega}$
and  $H \lesssim H_{\Omega}$, i.e., at ${\overline\Delta\sigma} > 0$
and ${\overline\Delta\sigma} < 0$~\cite{18,19}.  
As follows from Eq.~(14), $\Delta \rho_{xy}$ comprises
a component (corresponding to the  term proportional to
$\Delta\sigma_H \propto P\Phi_H$) which is   an even function
of the magnetic field. 
According to Eqs.~(13) and (14), 
this component essentially depends on the polarization
of microwave radiation.
The term in Eq.~(14)
dependent on $\Delta\sigma_D \propto P\Phi_D$ changes its sign when
the magnetic field reverses.
Figure~1 shows the dependence of the Hall microwave photoconductivity
$\Delta\sigma_H$ on  
the magnetic field normalized by $H_{\Omega} = mc\Omega/e$
calculated using the above equations.
Figure~2 shows the magnetic field
dependences of the dissipative and Hall resistivities 
in the vicinity of  the cyclotron resonance
calculated using general relationships between
the components of the conductivity and resistivity
tensors. It is assumed that the microwave power
slightly exceeds the threshold of ANC.
One can see that the $\Delta\rho_{xy}$ vs $H$ dependence, 
in contrast to the magnetic field dependence of  $\rho_{xx}$,
exhibits an antisymmetrical behavior in reasonable agreement
with the experimental results~\cite{18,19} (compare with Fig.~2(b)
from Ref.~\cite{18} and Fig.~1(c) from~Ref.~\cite{19}).
As seen from Fig.~2, the calculated dissipative resistivity
$\rho_{xx}$ is negative when $|H| \lesssim H_{\Omega}$. 
In this range of magnetic fields, uniform states of a 2DES
can be unstable resulting in the formation of complex domain structures
with the net resistance close to zero. 
In   these calculations we put
$\sigma_H/\sigma_D = 100$ at $H = 1.5$~kG and $
\gamma = \Gamma/\Omega = 0.15$. 
By convention
the plots shown in Figs.~1 and 2 can be attributed
to a AlGaAs/GaAs 2DES 
with the electron mobility $\mu = 1.5\times10^7$~cm$^2$/Vs
irradiated with microwaves with the frequency  $\Omega/2\pi = 50$~GHz.  
The dependence of $\rho_{xx}$ and $\Delta\rho_{xy}$
on the microwave power becomes nonlinear
at its  elevated values. This also leads to an alteration of the 
polarization dependences.  However,
this problem needs a separate study. 

In summary, we calculated the dependences of the dissipative and Hall
resistivities in 2DES under microwave radiation
on the magnetic field (its strength and sign)
and the radiation polarization. We showed that microwave irradiation
can lead to
variations of the Hall resistivity 
which are   even and odd oscillatory functions of the magnetic field.
The obtained  dependences are consistent with recent experimental
data~\cite{18,19}.

The author is grateful to K.~von~Klitzing for discussion
and to S.~Studenikin,
P.~D.~Ye, and I.~Aleiner for 
comments.

\end{document}